\documentclass[journal]{IEEEtran}

\usepackage{tikz,xcolor,hyperref}
\definecolor{lime}{HTML}{A6CE39}
\DeclareRobustCommand{\orcidicon}{%
 \begin{tikzpicture}
    \draw[lime, fill=lime] (0,0) 
    circle [radius=0.16] 
    node[white] {{\fontfamily{qag}\selectfont \tiny ID}};    \draw[white, fill=white] (-0.0625,0.095) 
    circle [radius=0.007];    \end{tikzpicture}
    \hspace{-2mm}}
\foreach \x in {A, ..., Z}{%
    \expandafter\xdef\csname orcid\x\endcsname{\noexpand\href{https://orcid.org/\csname orcidauthor\x\endcsname}{\noexpand\orcidicon}}
    }
\hypersetup{hidelinks}

\RequirePackage[normalem]{ulem} 
\RequirePackage{xcolor}
\usepackage{soul}
\providecommand{\DIFadd}[1]{{{{\protect\color{black} {#1}}}}}
\providecommand{\DIFdel}[1]{{ }} 

\usepackage{amsmath,amssymb,amsfonts}
\usepackage{graphicx}
\usepackage{cite}

\usepackage{algorithmic}
 
\usepackage{textcomp}
\usepackage{multirow}
\usepackage{tabularx}
\usepackage{hhline}
\usepackage{array}
\newcolumntype{P}[1]{>{\centering\arraybackslash}p{#1}}
\newcolumntype{L}[1]{>{\raggedright\let\newline\\\arraybackslash\hspace{0pt}}m{#1}}
\newcolumntype{C}[1]{>{\centering\let\newline\\\arraybackslash\hspace{0pt}}m{#1}}
\newcolumntype{R}[1]{>{\raggedleft\let\newline\\\arraybackslash\hspace{0pt}}m{#1}}
\usepackage{booktabs}

\usepackage[acronym,nomain]{glossaries}
\makeglossaries

\newacronym{cnn}{CNN}{Convolutional Neural Network}
\newacronym{gpu}{GPU}{Graphics Processing Unit}
\newacronym{rcnn}{RCNN}{Recurrent Convolutional Neural Network}
\newacronym{ivus}{IVUS}{Intra-Vascular Ultrasound}

\newacronym{nurd}{NURD}{Non-Uniform Rotational Distortion}

\newacronym{sd}{SD}{spectral-domain}
\newacronym{ekf}{EKF}{Extended Kalman Filter}
\newacronym{mse}{MSE}{Mean Squared Error}
\newacronym{std}{STD}{Standard Deviation}

\newacronym{rgb}{RGB}{Red, Green\& Blue}
\newacronym{bn}{BN}{Batch Normalization}
\newacronym{sgd}{SGD}{Stochastic Gradient Descent} 
\newacronym{oct}{OCT}{Optical Coherence Tomography} 

\newacronym{snr}{SNR}{Signal-to-Noise Ratio} 
\newacronym{gan}{GAN}{Generative Adversarial Networks} 
\newacronym{dof}{DoF}{Degree of Freedom} 

\newacronym{fov}{FoV}{Field of View} 

\ifCLASSINFOpdf
\else
\fi

\usepackage[switch]{lineno} 
\begin{document}
 
\title{Data Stream Stabilization for Optical Coherence Tomography Volumetric Scanning}

\author{Guiqiu~Liao\orcidA{}, 
        Oscar~Caravaca-Mora\orcidB{}, 
        Benoit~Rosa, 
        Philippe~Zanne, 
        Alexandre~Asch,
        Diego~Dall’Alba,
        Paolo~Fiorini,~\IEEEmembership{Fellow,~IEEE,}
        Michel~de~Mathelin,~\IEEEmembership{Senior member,~IEEE,}
        Florent Nageotte and
        Michalina~J. Gora 

\thanks{Manuscript received January 31st, 2021; revised May 12th, 2021, accepted June 15th, 2021.  This work was supported in part by the European Union's Horizon 2020 research and innovation programme under the Marie Sklodowska-Curie grant agreement No 813782 - project ATLAS.
}
\thanks{Guiqiu Liao is with ICube, UMR 7357 CNRS-University of Strabourg, Strasbourg, France, and also with Department of Computer Science, University of Verona, Verona, Italy (e-mail: liao.guiqiu@etu.unistra.fr). }
\thanks{Oscar Caravaca-Mora, Benoit Rosa, Philippe Zanne, Alexandre Asch, Michel de Mathelin, Florent Nageotte and Michalina J. Gora are with ICube,  UMR 7357 CNRS-University of Strabourg, Strasbourg, France (e-mail: caravacamora@kuleuven.be; b.rosa@unistra.fr; zanne.philippe@unistra.fr; alexasch3h@gmail.com; demathelin@unistra.fr; Nageotte@unistra.fr; gora@unistra.fr).}
\thanks{Diego Dall’Alba and Paolo Fiorini are with Department of Computer Science, University of Verona, Verona, Italy (e-mail: diego.dallalba@univr.it; paolo.fiorini@univr.it).}}

\markboth{IEEE Transactions on  Medical Robotics \& Bionics\,~Vol.~xx, No.~xx, June~2021}%
{Shell \MakeLowercase{\textit{et al.}}: Bare Demo of IEEEtran.cls for IEEE Journals}
%



\maketitle

\begin{abstract}
Optical Coherence Tomography (OCT) is an emerging medical imaging modality for luminal organ diagnosis. The non-constant rotation speed of optical components in the OCT catheter tip causes rotational distortion in OCT volumetric scanning. By improving the scanning process, this instability can be partially reduced. To further correct the rotational distortion in the OCT image, a volumetric data stabilization algorithm is proposed. The algorithm first estimates the Non-Uniform Rotational Distortion (NURD) for each B-scan by using a Convolutional Neural Network (CNN). A correlation map between two successive B-scans is computed and provided as input to the CNN. To solve the problem of accumulative error in iterative frame stream processing, we deploy an overall rotation estimation between reference orientation and actual OCT image orientation. We train the network with synthetic OCT videos by intentionally adding rotational distortion into real OCT images. As part of this article we discuss the proposed method in two different scanning modes: the first is a conventional pullback mode where the optical components move along the protection sheath, and the second is a self-designed scanning mode where the catheter is globally translated by using an external actuator. The efficiency and robustness of the proposed method are evaluated with synthetic scans as well as real scans under two scanning modes.
\end{abstract}

\begin{IEEEkeywords}
Optical Coherence Tomography, Video Stabilization, Non-uniform Rotational Distortion, Convolutional Neural Network.
\end{IEEEkeywords}

%
\IEEEpeerreviewmaketitle

\section{Introduction}
 
\IEEEPARstart{O}{ptical} Coherence Tomography (OCT) is an emerging medical imaging modality for diagnosing diseases in cardiovascular~\DIFadd{system}, airways and gastrointestinal tract \cite{b_OCT_R}.
The main~\DIFadd{challenges} related to these~\DIFadd{applications are} delivery and collection of light from the volumetric scanning of the area of interest. Fourier domain OCT in a single measurement provides a one-dimensional in-depth information and in order to collect three-dimensional information the optical beam~\DIFadd{has} to be scanned  over the tissue. Various designs of specialized micro-optical catheters have been developed to achieve this goal \cite{catheter}.

\begin{figure}[t!]
\centerline{\includegraphics[width=0.95\columnwidth]{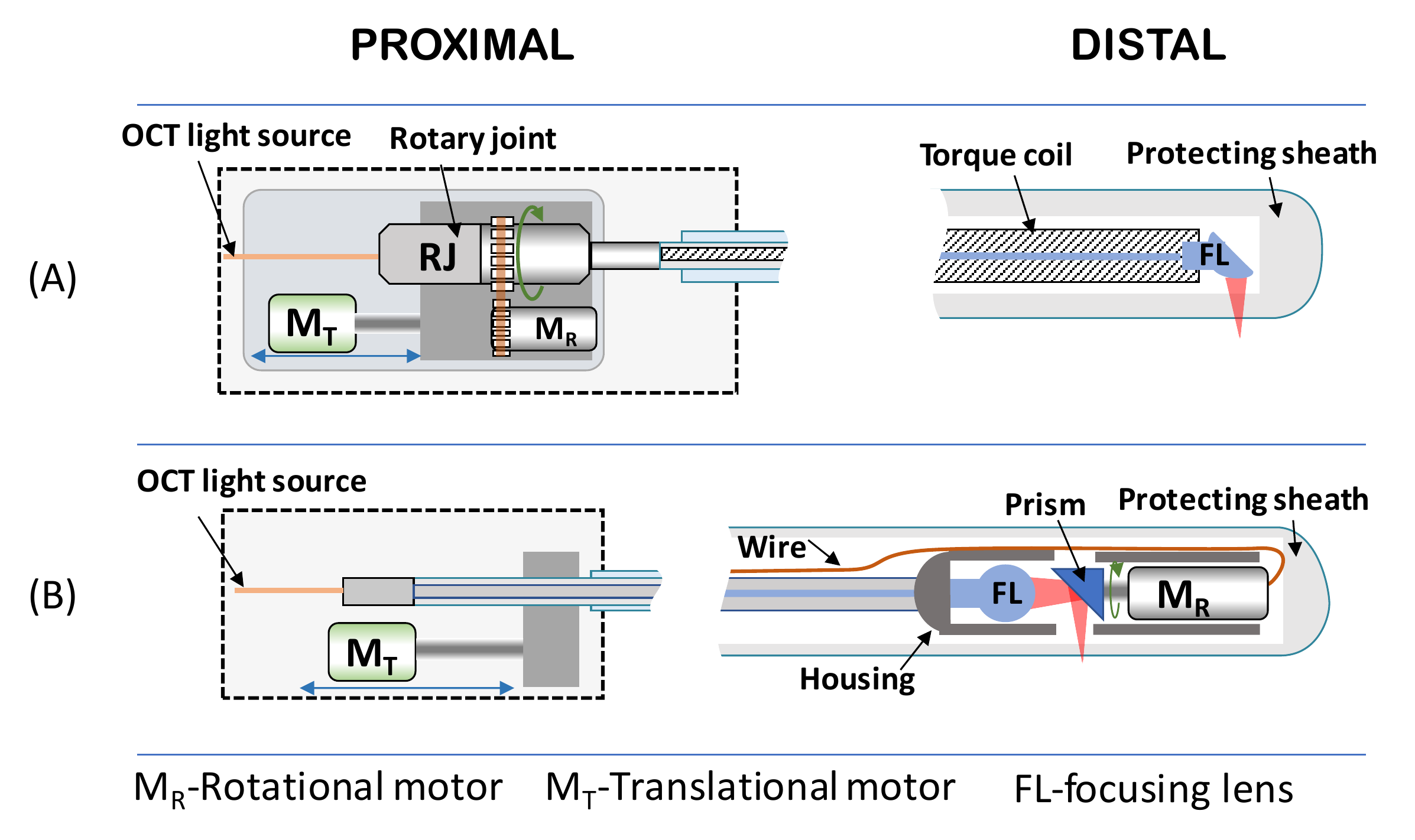}}
\caption{Schematics of (A) proximal-scanning with a fiber-optic rotary joint; (B) distal-scanning endoscope with a micro-motor.}
\label{fig_distal}
\end{figure}

A circumferential two-dimensional scan can be performed by rotation of an optical beam reflected to the side of the probe using a micro-motor on the distal tip, or by a proximal rotational actuation, which is remotely connected to the distal optical components with a torque coil (see Fig. \ref{fig_distal}).
Volumetric scanning, in both cases, is then typically effectuated by pulling back the rotating optical core to create a helical scan. This approach for volumetric scanning was originally developed for cardiovascular applications where an OCT catheter is inserted using a guidewire into the location of interest and a pullback needs to be synchronized with a blood occlusion \cite{b2_c}.
It was then adopted in gastrointestinal imaging both in low-profile and balloon catheters, which are inserted in the digestive system using \DIFadd{a} working channel of an endoscope \cite{balloon}.
Most recently, an internal pullback scanning system was also developed for a \DIFadd{tethered} capsule device \cite{capsule}.
A high precision short segment pullback enabled high quality en-face imaging that could not be achieved with a standard tethered capsule devices typically pulled back manually.

Even though motorized rotational and pullback actuation provides good control of the OCT beam position, scanning distortions are still very common \cite{balloon,capsule}. \gls{nurd} is crucial for catheter-based imaging systems such as endoscopic OCT but also intravascular ultrasound \cite{b1,b2}.

In proximal rotation devices \gls{nurd} can be observed due to mechanical friction between rotating optical components and the protecting sheath, mostly caused by bending of the catheter inserted in the tortuous path of the digestive, cardiovascular or respiratory system. Miniaturization of motors enabled distal scanning and more compact \cite{b_OCT_R} and more easily miniaturized designs\cite{b15}, in which instabilities are only related to the quality of the motor itself. However, these designs require electrical connection of the distally located motor, which adds additional safety risks to the device and \DIFadd{causes} partial shadowing of the tissue by the wires.
Non-linear transmission of displacement, and imperfect synchronization between data acquisition and scanning speed also causes distortions of the longitudinal scanning. This is typically observed as drift in the 3D reconstruction. 

We have previously investigated the NURD performance of 
an OCT catheter that was developed by our team for real-time assistance during minimally invasive robotized treatment of colorectal cancer \cite{oscar}.
This robotized OCT catheter can be inserted in one of the two instrument channels of the  robotized interventional endoscope and teleoperated using a master controller. In this article, we focus on studying the volumetric imaging instability of proximal scanning endoscopic OCT catheters, and on stabilizing the OCT image by proposing both a new robotic 3D scanning and a software approach. A new scanning method was motivated by redundancy of two degrees of freedom (DoF) in the OCT enhanced robotized interventional endoscope. The OCT catheter once inserted in the robot has two DoF for volumetric imaging (rotation and pullback of the inner optical core, provided by the OCT scanning system) and 3 DoF for maneuverability (rotation, translation and bending, provided by the robotics system \cite{stras}). As can be seen both rotational and translation are redundant, but OCT rotational scanning requires at least 3000 rpm, which cannot be achieved by a motorized instrument driver. The required translation speed of 2.5 mm/s, on the other hand, can be achieved by both inner core pullback or outer tool pullback. The outer pullback can reduce the instability of proximal scanning with a moving torque coil, but not perfectly, because the residual friction with the protecting sheath cannot be completely eliminated, and thus we propose a software correction to further decreased distortions. 
 
When using the information within OCT images, two challenges need to be addressed for the software \DIFadd{correction of the \gls{nurd}}. The first challenge is to compensate the \gls{nurd} on-line, which means that the algorithm should only take historical OCT data to stabilize the scanning video. This can allow the catheter to correctly access the cross-sectional  structure of tissues in real-time, which can dramatically reduce the time it takes for medical doctors/physicians to make a diagnostic decision. Unlike the off-line stabilization that uses the entire data after a scan to optimize the warping parameters, on-line \gls{nurd} correction typically follows an iterative processing pipeline where the estimation error accumulates. In this case, common iterative algorithms fail when a certain frame has a large stabilization error or when the processing time is long. Using optical markers is an alternative choice to stabilize rotational distortion, but it will block the OCT laser beam and degrade the image quality. \DIFadd{Thus} the second challenge \DIFadd{is that the algorithm should not use any type of additional objects to track the orientation shifts}. 

We formulate the problem as finding a warping vector for each frame, to shift each individual scanning A-line to its correct position. This can be done based on computing the cross-correlation or Euler distance between A-lines in the new uncorrected frame and in previous stabilized frames \cite{b12,b13,b14,b15}. However, in a real application it is hard to ensure that every A-line of OCT images can provide visible features for correlation because of the limited \gls{fov} of OCT catheters (part of the OCT image is filled with background noise when the tubular lumen is larger than the catheter \gls{fov}). Given the prior knowledge \cite{b12,b13} that the warping path should be continuous through the whole OCT frame stream,
the problem of searching the path in a computed correlation matrix is highly similar to a continuous boundary line prediction problem. Motivated by this, we seek to leverage the state of art machine learning techniques for the boundary line prediction method \cite{b33_bound,b20} as a solution to optimize the warping vector estimation.

Although the warping vector estimation accuracy can be improved, the accumulative drift still exists in long data stream processing. We address this issue by estimating an overall rotation value for each frame, and compensating the drift error of the \gls{nurd} warping vector using this overall rotation value. To do this, the algorithm exploits the consistency of the OCT protecting sheath as a cue for overall rotation estimation, and this overall rotation is a direct and robust estimation (without time-integration process), although it only provides a low accuracy because it relies on limited information. We deploy a fusion algorithm \cite{fusion3} to integrate the overall orientation and element-wise NURD estimation, which is inspired the fusion of gyroscope and accelerometry for attitude angle estimation \cite{fusion4}.

We demonstrate our method in two different 3D scanning modes: internal pullback scanning and outer tool scanning (see Fig. \ref{fig_internal} (A) and Fig. \ref{fig_internal} (B)). For the internal pullback scanning, we deploy an additional volumetric sheath registration. We train the CNN based \gls{nurd} estimation algorithm with semi-synthetic data by intentionally distorting OCT images. We evaluate the accuracy of the \gls{nurd} estimation with semi-synthetic data, and test its robustness on real data. An ablation study is
demonstrated by disabling parts of the full algorithm.

\section{Related Work}
In this section we provide an overview of previous research on \gls{nurd} correction for catheter-based imaging systems, followed by an introduction to the state-of-the-art video stabilization research and \gls{cnn} research for  white light camera, which inspired the proposed method for endoscopic OCT stabilization. 

\subsection{NURD Correction}
Several \gls{nurd} correction methods have been proposed for catheter-based imaging systems \cite{b1,b12,b13,b14,b15, b16,b19}.  A frequency analysis of the texture of the \gls{ivus} image was used in \cite{b1} to estimate the rotational velocity. Matching based algorithms by minimizing the Euler distance of warped neighboring frames \cite{b12,b13}, or by maximizing the cross-correlation \cite{b14,b15,b19} have been proposed to track the shift error of A-lines. Specifically, in intravascular OCT imaging, intra-vascular stents may serve as landmarks that help to capture rotational distortion in pullback volumetric scanning \cite{b18}. However, matching based methods using cross-correlation or Euler distance  typically require highly correlated images. Even with tractable features between successive frame pairs, it is difficult to guarantee that the estimation of \gls{nurd} is correct, which leads to iterative drift in long data streams processing. Adding structural markers on the OCT sheath provides an extrinsic reference for orientation change\cite{b13}, but the markers block the OCT light and thus remove information about the tissue.

\subsection{Data Fusion for Rotation Estimation}

Similar to iterative \gls{nurd} estimation, the estimation of attitude angle with integral gyroscope data also has the problem of drift \cite{fusion1,fusion2}. The integral drift of a gyroscope is usually compensated by another robust angle estimation from sensors such as accelerometers and magnetometers \cite{fusion4,fusion5,fusion6,fusion7}. While a gyroscope provides excellent information about rapid orientation changes, it only provides relative orientation changes that gradually drift with their lifetime and temperature. An accelerometer or magnetometer, on the other hand, has a direct measurement of orientation but with lower accuracy. Various classes of filters were demonstrated to fuse accurate rapid relative (indirect) measurements with less accurate direct measurements such as with the \gls{ekf} \cite{fusion7}, the complementary filter\cite{fusion3,fusion5,fusion6} and a gradient based filter \cite{fusion4}.  
In a similar way to the role of an accelerometer or a magnetometer, another additional overall rotation can be estimated and fused with the \gls{nurd} estimation, which compensates the accumulative error.


\subsection{Video Stabilization}

In the research field of image processing, \gls{cnn} and deep learning related approaches stand as state-of-the-art. For OCT image processing, \gls{cnn} has been applied to tissue layer segmentation \cite{b26,b27,b28}, classification \cite{b29} and cancer tissue identification \cite{b30}, but there is no CNN application for OCT stabilization. On the other hand, the literature in video stabilization is richer for white light cameras than for medical imaging systems (including the OCT). \DIFadd{We} seek to fill in the gap between the common computer vision research field and that of the OCT imaging, by relying on the \gls{cnn} to enhance the efficiency of OCT frame stream stabilization.

 
For white light camera video stabilization, there are two types of approaches to model the problem. One seeks to directly estimate the camera path (position), and the video stabilization can be considered as a camera path smoothing problem \cite{stab2}. This formulation aims to stabilize homographic distortion cause by camera shaking, and recently a deep learning based method has been developed to learn from data registered by a mechanical stabilizer \cite{stab3}, which shows greater efficiency than traditional algorithms. The other type of approach models the instability of the video (or frame stream) as an appearance change \cite{stab4}. This modeling methodology can be adapted to different imaging systems beyond the white light camera. To formulate the appearance change, features matching algorithms or optical flow \cite{flow1,flow2} can be used. A recent study uses deep learning techniques to estimate an optical flow field representing shift map of pixels in the video frames, and then applies another \gls{cnn} regression module to estimate a pixel-wise warping field from the optical flow field \cite{stab1}. Inspired by this, we deploy a \gls{cnn} to estimate the element-wise \gls{nurd} warping vector from a correlation map which roughly interprets the \gls{nurd} distortion of a single frame. 

\subsection{CNN for path searching}

An essential step of the \gls{nurd} estimation is to search a continuous optimal path with large correlation value from the correlation map. Solutions applied in previous OCT stabilization studies \cite{b14,b15,b19} are mainly based on graph searching \DIFadd{(GS)}, and rely on local features of gradients, maxima, textures and other prior information. This type of technique is also a traditional way of contour tracing \cite{contour_tradition}. Recently deep learning based contour tracing techniques \cite{contour_DL1,contour_DL2,contour_DL3,b33_bound} has been demonstrated to be faster and more robust than traditional methods.

The state-of-the-art deep learning models for pixel-wise segmentation are based on adaptations of convolutional networks that had originally been designed for image classification. 
To solve dense prediction problems such as semantic segmentation\DIFadd{,} which are structurally different from image classification, striding and dilated \gls{cnn} \cite{dilated} are proposed to  systematically aggregate multiscale
contextual information without losing resolution. 
Path detection can be achieved with a pixel-wise segmentation architecture (\emph{e.g.} predict a binary map where the path position and background pixels have different values). This is a high-cost approach\DIFadd{,} which usually adopts a U-shape \gls{cnn} \cite{contour_DL2,contour_DL3,b33_bound} using up-convolution layers\cite{b37_de_con,b37_up_con}. We deploy a CNN to predict a single vector representing path coordinates from the correlation map. Compared with pixel-wise prediction, the proposed architecture can be more efficient and no post processing is required. Moreover, it is easy to integrate the continuity prior knowledge into the network training, by an additional term in the loss function. Since we consider the situation when part of the image has no feature, a simple cascade layer architecture \cite{b35_vgg} is not sufficient for spacial exploitation. Therefore we use a parallel scheme \cite{b22_google} which controls striding of convolution to extract hierarchical abstract information from the correlation matrix. 
\DIFadd{Moreover, the proposed network has shallower layers  than VGG \cite{b35_vgg} or Resnet \cite{b36_resnet} and thus requires less inference time. Our tests and results indicate that this choice is sufficient for the NURD stabilization task.}

\section{Methods}

\subsection{OCT volumetric imaging}
To compare the conventional pullback and robotic 3D scanning~performance we use an OCT instrument with an outer diameter of 3.5 mm \cite{oscar}. The instrument is terminated at the distal end with a transparent sheath on the tip, which allows three-dimensional OCT imaging using an internal rotating side-focusing optical probe with two proximal external scanning actuators. The instrument is connected to an OCT imaging system built around the OCT Axsun engine, with a 1310 nm center wavelength swept source laser and 100 kHz A-line rate. The OCT catheter is compatible with an instrument channel of a robotized flexible interventional endoscope \cite{stras}. The distal end of the robotized endoscope can be bent in 2 orthogonal planes, translated and rotated. The OCT catheter can be translated, rotated and bent in one plane. In addition, the rotation and translation of the inner OCT probe can also be controlled by a servo system.

\begin{figure}[t!]
\centerline{\includegraphics[width=0.9\columnwidth]{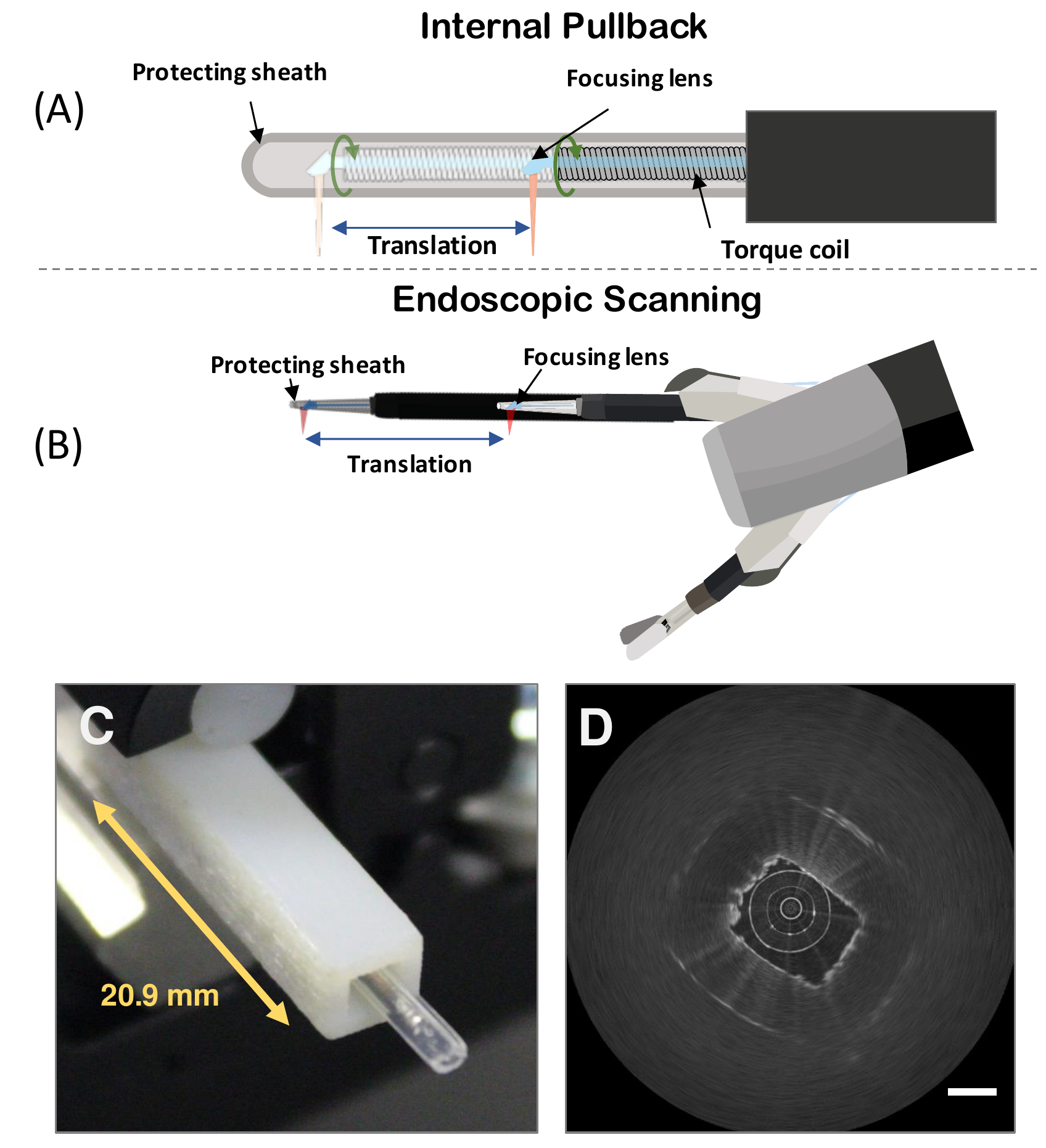}}
\caption{(A) Schematic of internal pull back for proximal-scanning OCT; (B) Schematic of robotic endoscopic scanning; (C) The rectangular test object; (D) An example of OCT B-scan, the scale bar corresponds to 1 mm.}
\label{fig_internal}
\end{figure}

We test two different types of volumetric scans with this system. One is an internal pullback scanning where the torque coil of the catheter moves through the protecting sheath, and another one is a robotic outer pullback of the whole tool where the torque coil delivers only rotational scanning.
Fig. \ref{fig_internal}
shows a schematic of the internal pullback scan and the outer tool scan.

\subsection{Data Stream Stabilization}

A \DIFadd{typical} OCT scanning system makes the assumption that the focusing lens at the distal tip rotates with a constant speed, so the OCT data acquisition system arranges A-lines of each frame with equal spacing to cover a 360 degrees region in the polar domain. 
However, for various reasons such as friction and non-constant motor speed the angular speed of the distal optical components may not be uniform over time, resulting in a distortion of the reconstructed B-scan in polar domain.
The OCT rotational distortion correction algorithm estimates a warping vector representing the angular position error of each independent A-line.
This vector is used to place A-lines at new positions in order to limit the NURD artifact.
\begin{figure}[t!]
\centerline{\includegraphics[width=1.0\columnwidth]{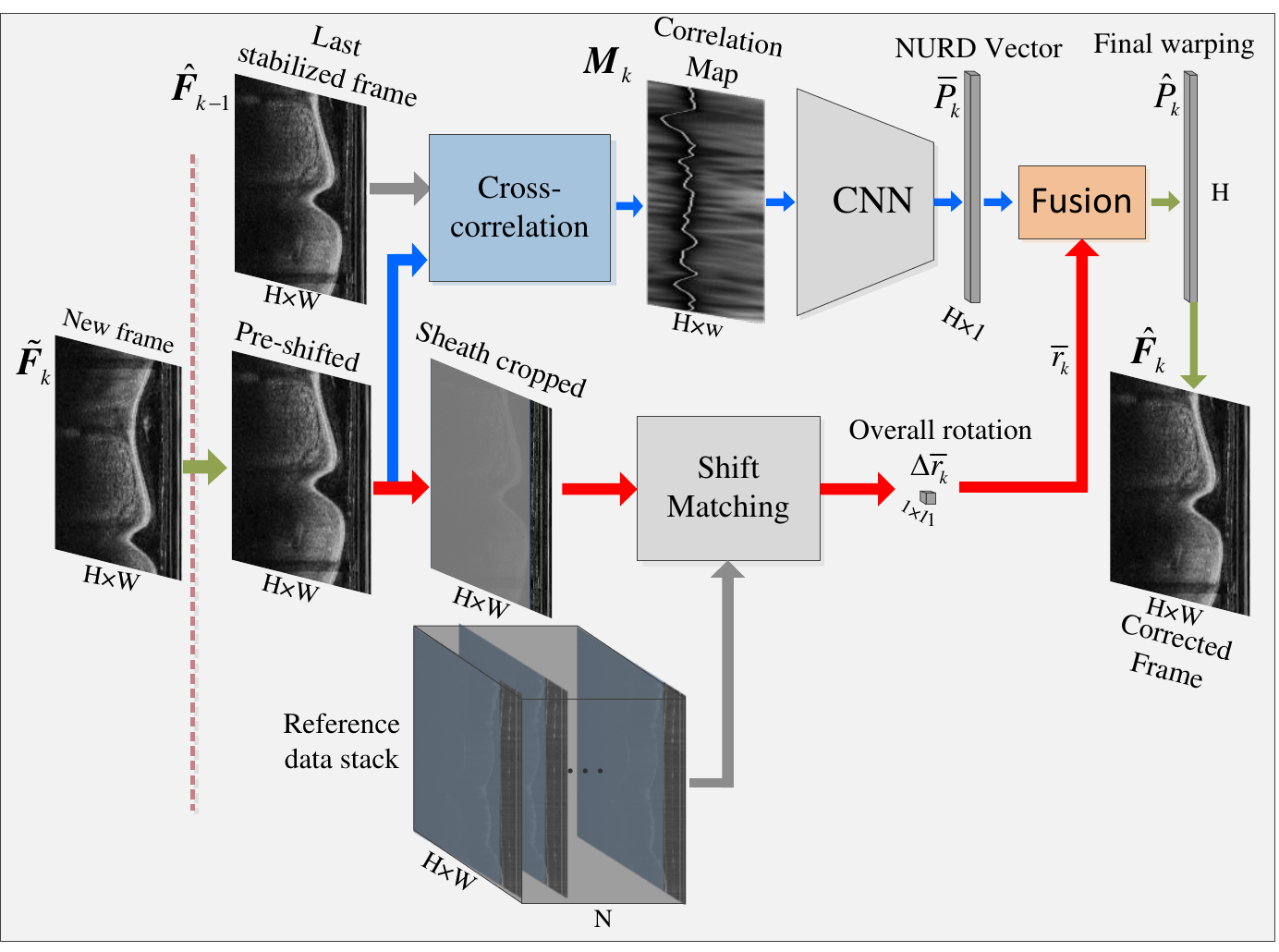}}
\caption{Algorithm pipeline of the proposed OCT stabilization method. The original new frame is first pre-shifted based on the previous estimated overall rotation. \DIFadd{Then the pre-shifted OCT image follows two paths in parallel to estimate two levels of artifacts. In the first path (with blue color) a correlation map between the pre-shifted frame and last stabilized frame is computed, and a CNN estimates an A-line level NURD vector. In the second path (with red color), the sheath image of the pre-shifted frame is used to match the reference sheath stack to estimate an overall rotation value. The final warping vector is a fusion of the A-line level NURD vector (from the first path) and the single rotation value (from the second path), which is used to shift each individual A-line of the new frame to obtain a new corrected frame}.    
}
\label{fig_method}
\end{figure}

The architecture of the proposed stabilization algorithm is shown in Fig. \ref{fig_method}, and is composed~\DIFadd{of} two parts. The first part iteratively estimates the NURD vector ${{\bar{P}}_{k}}$ with the last stabilized frame ${{\hat{\boldsymbol{F}}}_{k-1}}$ and the new frame ${{\tilde{\boldsymbol{F}}}_{k}}$, which is pre-shifted with a previous estimated overall rotation ${{\bar{r}}_{k-1}}$. The second part estimates the new overall rotation ${{\bar{r}}_{k}}$ with the pre-shifted ${{\tilde{\boldsymbol{F}}}_{k}}$ and a reference frame stack. 
The NURD warping vector ${{\bar{P}}_{k}}$ is fused with ${{\bar{r}}_{k}}$, to compensate the iterative drift of \gls{nurd} estimation. By doing so, a final warping vector ${{\hat{P}}_{k}}$ is estimated. ${{\hat{P}}_{k}}$ is equal in length to the polar image height $H$, which is the number of A-lines in a single OCT frame.  
Details on these two parts and the fusion algorithm are presented in the following two subsections.

\subsubsection{NURD estimation}
The first part of the stabilization algorithm follows two steps to estimate the \gls{nurd} vector between the previous stabilized ${{\hat{\boldsymbol{F}}}_{k-1}}$ and pre-shifted frame ${{\tilde{\boldsymbol{F}}}_{k}}$: First a correlation map ${{\boldsymbol{M}_{k}}}$ is calculated using cross-correlation, and then a CNN estimates the \gls{nurd} vector taking the ${{\boldsymbol{M}_{k}}}$ as input. 

Given two consecutive OCT frames with height $H$ (with $H$ A-lines in one frame) and width $W$ in polar coordinates, we used the Pearson correlation coefficient to reflect the similarity between two local image patches from these two frames:

\begin{equation} 
{{s}_{i,j}}=\frac{\sum\nolimits_{l=1}^{n}{{{\boldsymbol{f}}_{i,l}}{{\boldsymbol{f}}_{j,l}'}-n{{{\bar{\boldsymbol{f}}}}_{i}}{\bar{{{\boldsymbol{f}}}_{j}'}}}}{\sqrt{\sum\nolimits_{l=1}^{n}{\boldsymbol{f}_{i,l}^{2}-n\bar{\boldsymbol{f}}_{i}^{2}}}\sqrt{\sum\nolimits_{l=1}^{n}{{\boldsymbol{f}}_{j,l}'^{2}-n{\bar{\boldsymbol{f}_{j}'}}^{2}}}}
\label{eq_corr}\end{equation}

where  $\boldsymbol{f}_i \in {{\mathbb{R}}^{h\times W}}$ ($h \ll H$) is a local image patch centered at index position $i,i\in [0,H)$ of the newest frame. Correspondingly, ${{{\boldsymbol{f}}}_{j}'} \in {{\mathbb{R}}^{h\times W}}$ is a local image patch centered at index position $j$ of ${{\hat{\boldsymbol{F}}}_{k-1}}$ ( $j\in [i-w/2,i+w/2)$, $w$ is is the shifting window width). ${s}_{i,j}$ is one element of the correlation map ${{\boldsymbol{M}_{k}}} \in {{\mathbb{R}}^{H\times w}}$. The pixel index $l$ will operate through the patch size $n=h\times W$. $\bar{\boldsymbol{f}_{i}}$ and $\bar{\boldsymbol{f}_{j}'}$ are the mean values of patches ${\boldsymbol{f}_{i}}$ and ${\boldsymbol{f}_{j}'}$ respectively.
 
\begin{figure}[t!]
\centerline{\includegraphics[width=0.8\columnwidth]{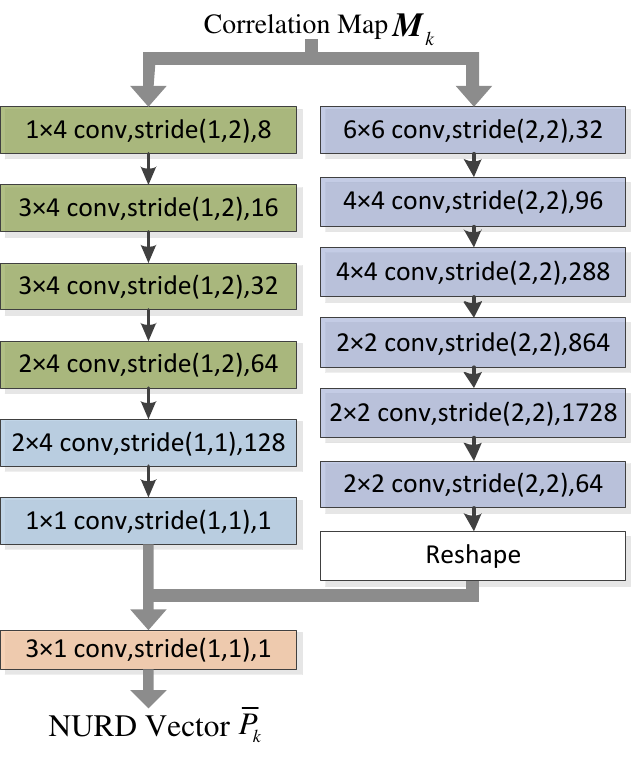}}
\caption{ Schematics of the CNN for NURD estimation. This network has two sub-branches that have different convolution kernels and striding.}
\label{fig_nets}
\end{figure}

The correlation map ${{\boldsymbol{M}_{k}}}$ contains information on the \gls{nurd}, and we design a CNN to further extract the \gls{nurd} vector. The detailed diagram of the \gls{nurd} estimating nets is shown in fig. \ref{fig_nets}, where each block gives the convolutional kernel size, the stride step length and the output feature map depth of each convolution layer.
Two convolution sub-branches are operated in parallel to
capture features and produce hierarchically coarse-to-fine responses. The main difference between these two sub-branches is the strides control.  
Both the left sub-branch and the right sub-branch of the \gls{nurd} estimation nets have 6 convolutional layers, and a LeakyReLU activation \cite{b34_relu} is used after each convolution layer. 

The left sub-branch always keeps the vertical stride as 1. 
The objective of this design is to emphasize the vertical spacial correspondence.
By doing so, the front 5 feature extraction layers can gradually reduce the feature map width from $w$ to 1, while maintaining the feature map height $H$ as the input's height. The depth of each convolution operation's output is twice as deep as its input (here we set the output depth of the first layer as 8). The $5th$ feature map $\boldsymbol{A}_{F}^{5}\in {{\mathbb{R}}^{H\times 1\times 128}}$ extracts 128 local features, which could include the minimal value position, edge, and boundary position. A final layer with kernel size $1\times 1$ and channel depth 128, reorganizes the $5th$ feature map and decrease channels to a sub-branch output of $\bar{P}$ with size $H\times 1$.

Considering that in certain situations ${{\boldsymbol{M}_{k}}}$ will miss valid information for some row $m_i$ when there is no feature in a patch (window) $\boldsymbol{f}_i$ of ${{\tilde{\boldsymbol{F}}}_{k}}$, the right sub-branch is added. This sub-branch loosens the horizontal stride to 2 to involve more spacial information. This is a more common architecture for many deep learning applications \cite{b35_vgg}. Compared to the left sub-branch, this sub-branch can extract higher abstract features which are less sensitive to local missing information in the correlation map. \DIFadd{We put larger kernels in the front layers to involve more common spatial information for neighboring positions of convolution, which prevents the under-fitting of feature extraction \cite{b_20_a}. The kernel size is reduced in deeper layers to ensure a good trade-off between kernel size and computational burden.}  
After the six convolutions, this subbranch outputs  a matrix $\bar{\boldsymbol{P}}_m''$ of size $(H/2^6)\times w$. To get an output of warping vector size $H\times 1$, a reshape operation is applied to the output of this sub-branch (\emph{i.e.} by sequentially connecting each $1\times w$ row). A final convolution layer with kernel size $3\times1$, input depth 2 and output depth 1,  combines the prediction of two sub-branches to get an estimation of $\bar{P}$.
The loss function for training the \gls{nurd} estimating nets uses the conventional $L_2$ loss and an additional continuity loss. A standard $L_2$ loss is described by Eq. \eqref{eq_L2}:

\begin{equation} 
{{L}_{2}}=\frac{1}{{{n}_{p}}}{{\sum\nolimits_{i=1}^{{{n}_{p}}}{({{P}_{i}}-{{{\bar{P}}}_{i}})}}^{2}}
\label{eq_L2}\end{equation}
where $P_i$ is the element of $P$, the true \gls{nurd} vector (ground truth), $n_p =H$ is the vector length. The $L_2$ loss function is commonly used for value estimation, while for this estimation task,
to take into account the prior knowledge on the continuity of the distortion vector \cite{b12,b13,b14}, we add the continuity loss as follows:

\begin{equation} 
L_c=\frac{1}{{{n}_{p}}-1}{{\sum\nolimits_{i=1}^{{{n}_{p}-1}}{({{\bar{P}}_{k,i}}-{{{\bar{P}}}_{k,i+1}})}}^{2}}
\label{eq_LC}\end{equation}

By calculating $L_c$, and combining it with $L_2$ in the network training, the attraction towards local minima with discontinuous \gls{nurd} estimation will be suppressed. The final loss for branch (A) is:

\begin{equation} 
L_A = \alpha L_c + (1- \alpha)L_2 
\label{eq_LA}\end{equation}
\subsubsection{Drift compensation}

It is possible to correct a short frame stack with just the \gls{nurd} estimation part presented in the previous subsection.
To compensate the accumulative error through iterations (especially for a long data stream), we estimate an overall rotation value ${{\bar{r}}_{k}}$. 

In conventional pullback scans only the sheath data is consistent no matter what the environment outside the sheath is. According to this, the overall rotation can be observed by matching real-time sheath images with pre-recorded reference sheath images. However, when using the unstabilized pullback scanning to record the reference images, it still suffers from rotational distortion. To ensure \DIFadd{that} the orientations of reference frames are correct, we follow a calibration procedure.
 The setup of the sheath registration is shown in Fig. \ref{fig_sheath} (A), which relies on a external calibration object (a straight and flat ruler). As shown in Fig. \ref{fig_sheath} (B), the raw reference data still has the rotational distortion, which leads to both the ruler and sheath images located at wrong direction. 
 We extract the contour surface of the ruler, and align the raw reference frame stack by minimizing the surface distance of all frames. By doing so, the rational error of the raw reference volumetric data is reduced from 59.4$^{\circ}$ to 2.79$^{\circ}$ (see Fig. \ref{fig_sheath} (C)). 
This calibrated reference data composes one of the inputs of the overall rotation estimation.
Another input is composed by real-time B-scans, where image outside the sheath is masked out and only the sheath part is used.

With a recorded sheath image stack $\boldsymbol S \in {{\mathbb{R}}^{H\times W\times N}}$ (N is the number of frames in the entire reference stack), we compute Euler distances to estimate the rotation deviation $\Delta \bar{r}_k$ of pre-shifted image ${{\tilde{\boldsymbol{F}}}_{k}}$ (pre-shifted by $\bar{r}_{k-1}$):
\begin{figure}[t!]
\centerline{\includegraphics[width=0.8\columnwidth]{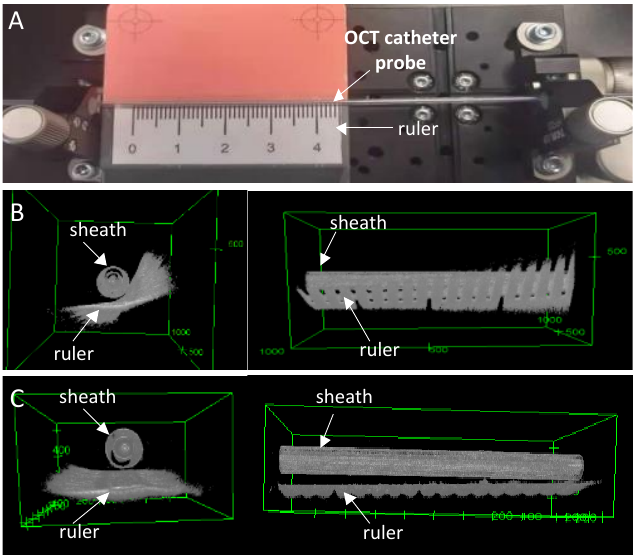}}
\caption{(A) Setup for sheath registration; (B) Original registration data in 3D view; (C) Calibrated registration data in 3D view.}
\label{fig_sheath}
\end{figure}

\begin{equation} 
{{d}_{r}}= \mathcal{L}({{\tilde{S}}_{k}}, S_k)
\label{eq_shift}\end{equation}
where ${{\tilde{S}}_{k}} \in {{\mathbb{R}}^{H\times W\times l}}$ is buffered pre-shifted original images, $S_k \in {{\mathbb{R}}^{H\times W\times l}}$ is the corresponding selected buffer indexed at $k$ of $\boldsymbol S$ ($l \ll N$, and assuming that in every pullback scanning the torque coil moves with equal 
interval). The function $\mathcal{L}$ calculates Euler distance of two input buffers. For the pullback scan, $S_k$ is the corresponding reference image in the data stack, while to correct an outer robotic tool pullback 3D scan the algorithm can be simplified and just~\DIFadd{use} the first frame ${{\boldsymbol{F}}}_{0}$ as the reference, because in this scanning condition the relative longitude position between sheath and rotating ball-lens is constant. For both ${{\tilde{S}}_{k}}$ and $S_k$ only the sheath part is used for distance calculation. The index $r \in (-w,w)$ shifts within a given window $w$ to calculate the distance $d_r$, and the position with the lowest distance value is considered as the matched angular position of $\Delta \bar{r}_k$. Finally, the overall rotation used to compensate the \gls{nurd} drift is computed by $\bar{r}_{k}=\bar{r}_{k-1} + \Delta \bar{r}_k$.
 
We use the concept of a \emph{PI Complementary Filter} \cite{fusion5} to fuse the $\bar{P}_k$ vector with the $\bar{r}_k$ value, which is shown to be efficient and able to solve the warping vector integral drift problem.
A discrete form of PI complementary filter is:

\begin{equation} 
{{\hat{P}}_{k}}={{k}_{p}}{\bar{P}_{k}}+(1-{{k}_{p}}){{\bar{r}}_{k}}\boldsymbol{1}+{{k}_{i}}{{I}_{k}}
\label{eq_pi2}\end{equation}
\begin{equation} 
{{I}_{k}}={{I}_{k-1}}+({{\bar{r}}_{k}}\boldsymbol{1}-{{\hat{P}}_{k}})
\label{eq_ki}\end{equation}
where $k_p$ and $k_i$ are PI compensating gains. ${{I}_{k}}$ is the integral component vector. $\boldsymbol{1}$ is a vector of ones.  Each element ${{\hat{P}}_{k,i}}$ of $\hat{P_k}$ represents the angular distance between the position of the $i$th A-line of ${{\tilde{\boldsymbol{F}}}_{k}}$ and its correct position in polar domain. Applying $\hat{P_k}$ to this raw frame ${{\tilde{\boldsymbol{F}}}_{k}}$, a stabilized frame ${{\hat{\boldsymbol{F}}}_{k}}$ is obtained.

\subsection{Experiment}
\subsubsection{Data}
We test the proposed stabilization algorithm for both the internal conventional pullback and \DIFadd{the} outer scanning using a custom made OCT system (see description in section III.A). We~collect videos with the testing object (fig. \ref{fig_internal} (C)) to evaluate the performance for both~types of scanning mode, since it has a special geometry. 6 scans were collected for each scanning mode, and each data stream comprises 500 frames. For the internal pullback, one reference scan is collected following the procedure described in section III.B.(2), and this reference scan is used in the correction of all internal pullback scans. To perform quantitative analysis, we also use synthetic videos generated from a variety of OCT images  \cite{b41_data1,b41_data2,b41_data3} to test the proposed algorithm, since it is almost impossible to manually annotate the \gls{nurd} in the OCT data. We used the published videos from \cite{b41_data1,b41_data2,b41_data3}, and extracted all the OCT images out and transformed them into the same resolution in the polar domain. Approximately 5000 images which cover the catheter based OCT imaging in intravascular, digestive tract and lung respiration airway are obtained from these public videos.

\subsubsection{NURD estimation networks training}

The network is only trained with synthetic videos by intentionally adding \gls{nurd} artifacts to the \gls{oct} images, and tested on both true \gls{oct} scans and synthetic scanning videos. 
\DIFadd{
To ensure that the artificially added distortion covers the distribution of artifacts present in real videos, we first compute the correlation map $\boldsymbol{M}_k$ of every neighboring raw image pairs in real videos, and then measure the maximum NURD error with a graph path searching method [11]. The range of the GS-observed NURD was then expanded by 1/3 on each side. For each synthetic image, the synthetic NURD vector was then randomly generated within this expanded range. In addition, we implemented data augmentation that included geometric transformations, noise addition, brightness/contrast modification and OCT speckle/shadow simulation. By doing so the domain of artifacts of synthetic videos should cover the distribution of real artifacts.}
We randomly split all the collected images by 1:1:1 into train, validation and test data. For both the \gls{cnn} training and algorithm deployment, we use the same PC with an Nvidia Qt1000 graphic card and Intel i5-9400H CPU. The code is implemented using the Pytorch framework\cite{b40pytorch}. For training, Batch Normalization (BN) is used right after convolution and before activation \cite{b_bn}, dropout is not used \cite{b_dpout} and weight initialization is performed following the method described in \cite{b_ini}.
 The CNN is trained with the \gls{sgd} weights optimization method (we used a weight decay of 0.0001 and a momentum of 0.9). 

\subsubsection{Metrics}
 
For synthetic videos we calculate the \gls{nurd} estimation error to evaluate the accuracy. For true videos, no guaranteed ground truth is available. We calculate the normalized \gls{std} $\sigma$ for a certain time window of videos to access the stabilization performance \cite{b12}. The definition of \gls{std} is:

\begin{equation} 
\sigma = \frac{1}{N_{sig}}\sum\nolimits_{i=1,j=1}^{N_{sig}}{\bar{\sigma}(f_{i,j})}
\label{eq_std}\end{equation}

where ${\bar{\sigma}(f_{i,j})}$ is the \gls{std} calculated with pixel $f_{i,j}$ in one stacked frame stream, $i$ and $j$ are selected pixel indices on horizontal and vertical axis respectively. $N_{sig}$ is the number of pixels used to calculate $\bar{\sigma}$. 
Based on the STD calculation, we give a threshold to differentiate unstable pixels from stable pixels, and count the number of unstable pixels for each frame.

\begin{figure*}[!thb]
\centerline{\includegraphics[width=\textwidth]{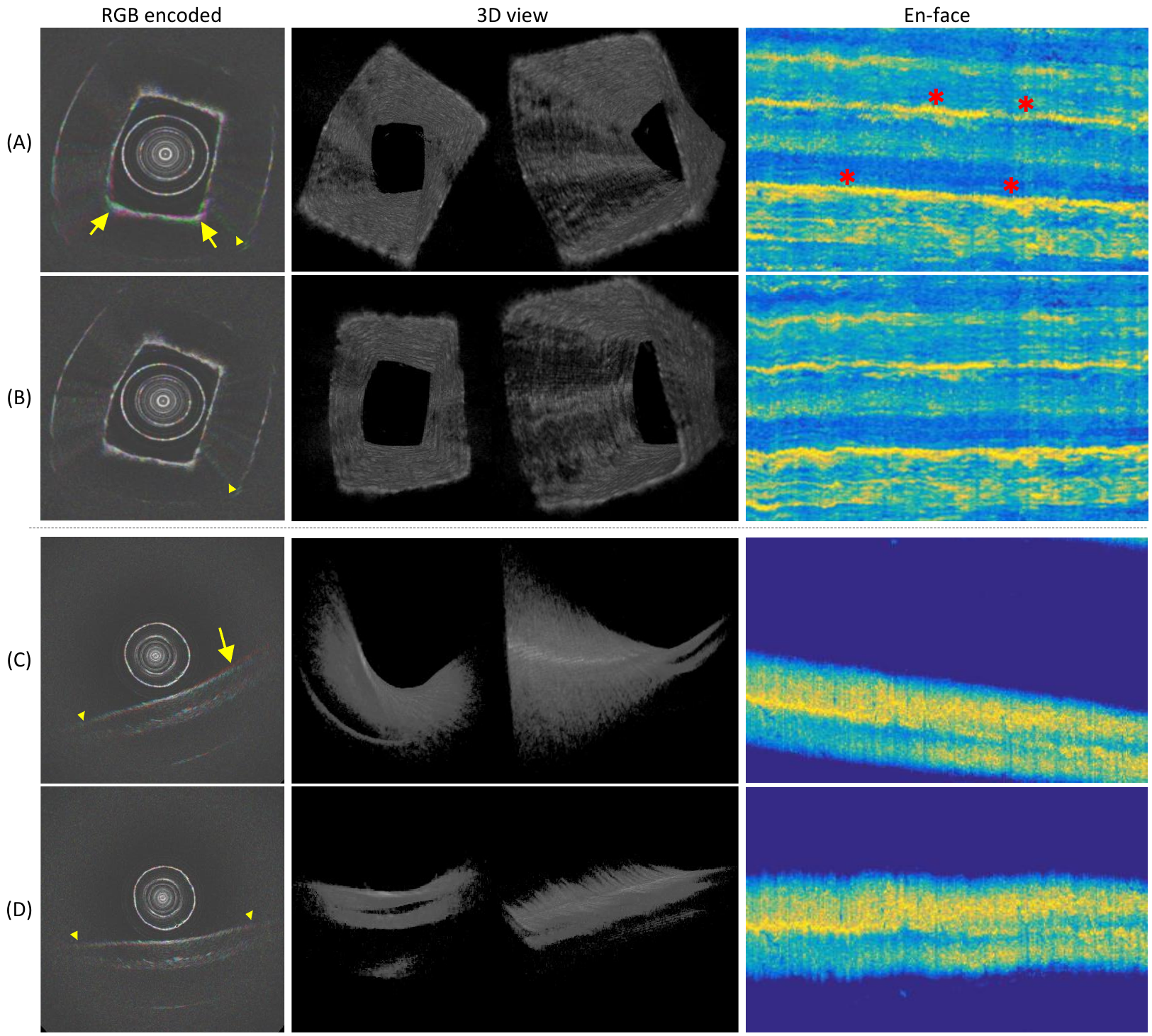}}
\caption{Result presented in RGB encoded image, 3D view and en-face projection. (A) shows results of an original scan of rectangular phantom and (B) shows the corresponding stabilized data. (C) shows results of an original scan of layered colon tissue phantom and (D) shows the corresponding stabilized data. \DIFadd{The yellow arrows point out rotation artifacts represented by colorful pixels in B-scans, the yellow arrowheads point out colorful pixels caused by the features changes of the object itself, and the red asterisks mark out aliasing artifacts in en-face projections.}}
\label{fig_3d_result}
\end{figure*}

To evaluate the rotation error through a complete scan, we compute en-face projections \cite{b14,b15} of the OCT videos where each A-line is accumulated to one single value, so that the volumetric data in the polar domain are projected into 2D images. In this case the vertical Y axis corresponds to a circumferential scanning (B-Scan) and the horizontal X axis to a longitudinal volumetric scanning (3D Scan). By counting the shift of max intensity pixels in a whole stack, we can evaluate the precession angle. We also measure the local fluctuation on the en-face image\DIFadd{,} which represents the local shaking between neighboring frames. 

\section{Results}

\subsection{Pullback actuation comparison}

To compare~\DIFadd{distortions}~\DIFadd{we effectuated} two \DIFadd{proposed} types of scans within a 21 mm long rectangular tube target 3D printed in polyjet VeroWhite material (Stratasys,  Rehovot, Israel), as shown in Fig. \ref{fig_internal} (C).
Table \ref{tab_internal_vs_robot} summarizes~\DIFadd{statistical analysis of distortion} of these 2 types of scans. We have also compared them to \DIFadd{a} stationary scanning without moving the optical components on the catheter tip, which is used as a baseline for hardware instability. It can be seen that the stationary scan data has lower precession, local rotation angle and \gls{std} (we also separate the sheath image STD from the full image STD) compared to any of the volumetric scanning.

\begin{table}[!t]
\caption{ Statistics of different scanning methods}
\label{tab_internal_vs_robot}
\centering
\setlength\arrayrulewidth{0.6pt}
\setlength\doublerulesep{0.8pt} 
\begin{tabular}{C{0.2\columnwidth}C{0.1\columnwidth}C{0.15\columnwidth}C{0.15\columnwidth}C{0.15\columnwidth}}
\toprule[0.8pt] \midrule[0.8pt]
 Method   & Preces- sion($^\circ$)$\downarrow$ &  local ($^\circ$)$\downarrow$ & STD$\downarrow$ & Sheath STD$\downarrow$ \\\toprule[0.9pt]
Stationary        & \textbf{11.6}       &  12.9$\pm$2.5   &  16.6$\pm$2.2             & \textbf{13.3$\pm$2.5}         \\
Conventional          & 79.6        &13.6$\pm$3.1  & 18.6$\pm$2.4        & 19.8$\pm$2.6      \\
Robotic 3D        & 24.7       & 12.3$\pm$1.9  & 17.2$\pm$2.9    & 16.9$\pm$2.6  \\\toprule[1.5pt]
\end{tabular}
\end{table}

These two volumetric scanning methods introduce instability into the data, and in particular the conventional internal pullback scanning presents larger image rotation between successive 2D radial frames, with a maximum precession angle of 79.6°. When performing volumetric robotic scanning of the OCT tool the image rotation (precession) is reduced to within the range of 24.6°. This constitutes an improvement and reduces the difficulty of software stabilization (see section III.B). Other metrics such as the local rotation and STD reflect the short term instability. Although the 3 scanning methods share similar short term instability, the stationary scan is sightly better than the internal pullback scan and robotic endoscopic scan. 


\subsection{Software Improvement}
\subsubsection{Quantitative results}
The proposed stabilization algorithm is applied to both scanning modes. A pixel-wise stack smoothing is deployed to be the baseline method (average filtering of each pixel using 3 consecutive frames). 

Table \ref{tab_baseline} shows the precession of the whole data stream, local fluctuation angle  and \textbf{mean pixel counts (MPC)} of unstable pixels. For both scanning modes, the precession angle is reduced by 75\%-80\%, and for the robotic scan the precession is stabilised to only 6$^\circ$. The baseline method can reduce the intensity derivation by blurring the 3D data, but does not improve the rotations and angular fluctuations. The proposed method not only reduces the overall rotations, but also reduces the pixel-wise instability without blurring the image or changing other information within every OCT frame (see the images in the subsection for qualitative results).
\begin{table}[!t]
\caption{ Comparison with baseline method }
\label{tab_baseline}
\centering
\setlength\arrayrulewidth{0.6pt}
\setlength\doublerulesep{0.8pt} 
\begin{tabular}{llll}
\toprule[0.8pt] \midrule[0.8pt]
Method              & Prec($^\circ$) $\downarrow$&local ($^\circ$)$\downarrow$& MPC($\times 10^3$)$\downarrow$  \\\toprule[0.9pt]
Original (conventional)  & 79.6       & 13.6$\pm$3.1   & 8.72$\pm$4.46 \\
Baseline (conventional)  & 79.1     &   13.1$\pm$3.2   &  2.94$\pm$2.31    \\
Proposed (conventional)      & 16.0      &  \textbf{2.61$\pm$0.4}   & \textbf{1.61$\pm$0.72}      \\ \toprule[0.9pt]
Original (robotic)    & 24.7         & 12.3$\pm$1.9     & 10.9$\pm$5.90    \\
Baseline (robotic)  & 24.6          & 12.1$\pm$2.3     & 3.79$\pm$2.83     \\
Proposed (robotic)     & \textbf{6.05}  & \textbf{3.02$\pm$0.8}  & \textbf{2.76$\pm$1.03}  \\ \toprule[1.5pt]
\end{tabular}
\end{table}

\subsubsection{Ablation Study}
\begin{table}[!t]
\caption{ ablation study for two scanning modes and synthetic scans}
\label{tab_ablatation}
\centering
\setlength\arrayrulewidth{0.6pt}
\setlength\doublerulesep{0.8pt} 
\begin{tabular}{lll}
\toprule[0.8pt] \midrule[0.8pt]
Method              & Prec($^\circ$) $\downarrow$ & MPC($\times 10^3$) $\downarrow$ \\\toprule[0.9pt]
Proposed full (conventional)  & \textbf{16.0}      & \textbf{1.61$\pm$0.72}  \\
w/o overall (conventional)  & 21.6        &  2.28$\pm$0.95    \\
w/o NURD  (conventional)      & 23.4        &  6.86$\pm$4.32     \\ \toprule[0.9pt]
Proposed full (robotic)  & \textbf{6.05}     & \textbf{2.67$\pm$1.03}  \\
w/o overall (robotic)  & 61.4       &  6.02$\pm$3.22    \\
w/o NURD  (robotic)      & 13.4         & 8.91$\pm$4.83      \\ \toprule[0.9pt]
original (synthetic)  & 132     &  19.8$\pm$5.30   \\
Proposed full (synthetic)  & \textbf{2.61}    & \textbf{2.60$\pm$1.43}  \\
w/o overall (synthetic)  &  136      &  3.01$\pm$1.72    \\
w/o NURD  (synthetic)      & 12.9        &  5.45$\pm$2.95      \\ \toprule[1.5pt]
\end{tabular}
\end{table}

\begin{figure}[tbh!]
\centerline{\includegraphics[width=0.95\columnwidth]{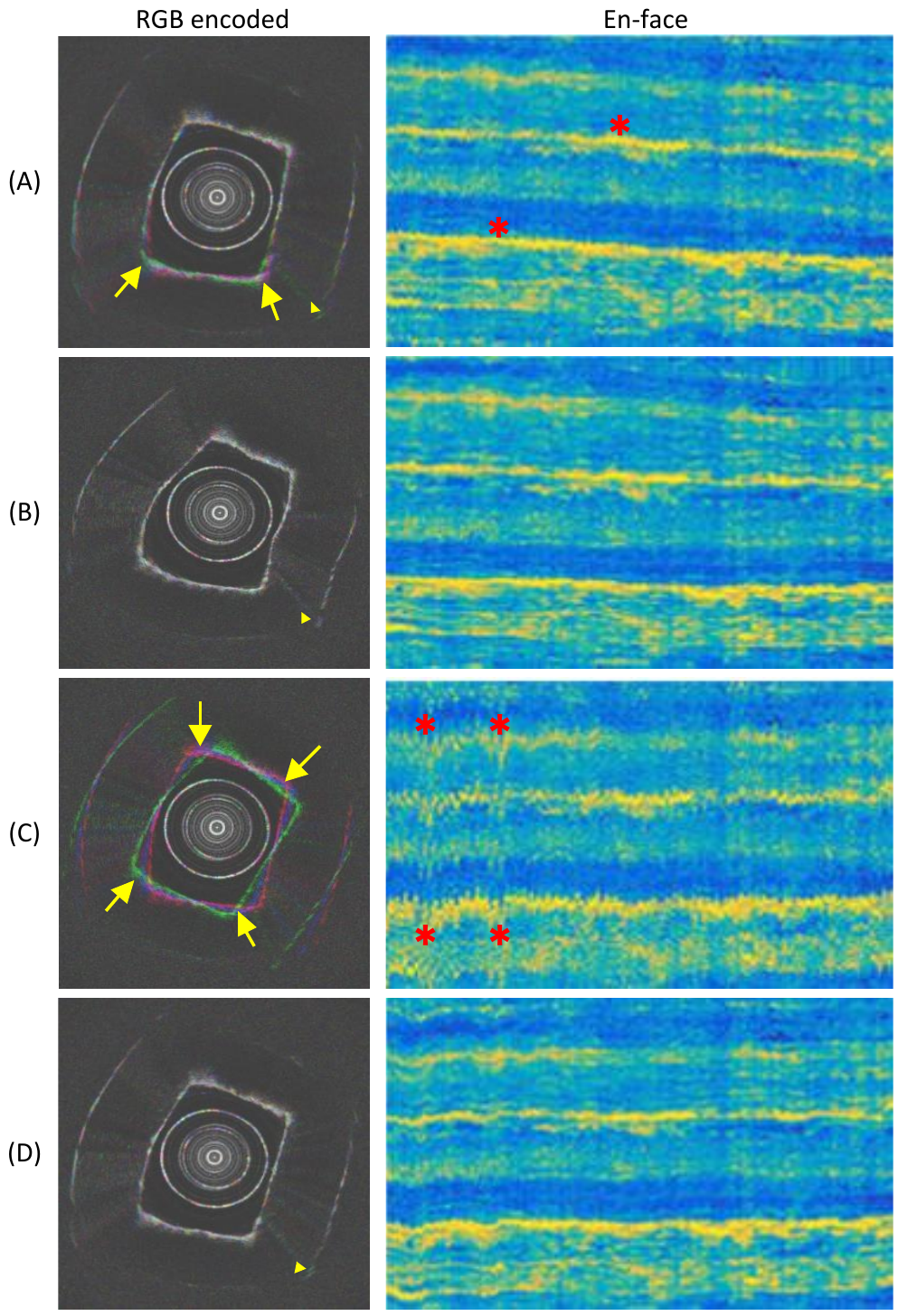}}
\caption{Ablation comparison results for scans of rectangular phantom. RGB encoded images and en-face projections of (A) original data, (B) stabilized data without (w/o) overall rotation estimation, (C) stabilized data with only overall rotation estimation and (D) stabilized data of the full proposed algorithm. \DIFadd{The yellow arrows point out rotation artifacts represented by colorful pixels in B-scans, the yellow arrowheads point out colorful pixels caused by the feature changing of the object itself, and the red asterisks mark out aliasing artifacts in en-face projections.}}
\label{fig_abla1}
\end{figure}
\begin{figure}[tbh!]
\centerline{\includegraphics[width=0.95\columnwidth]{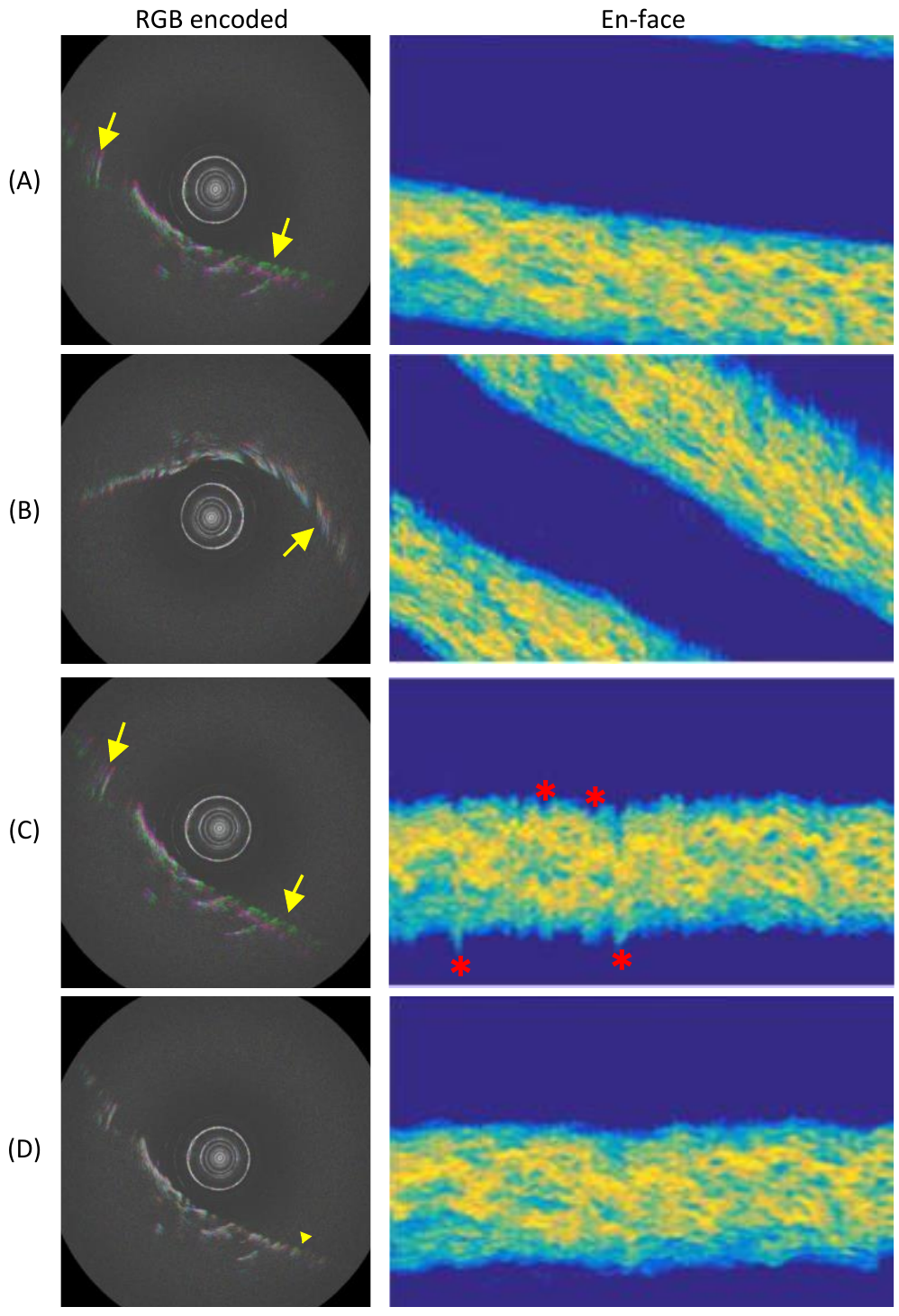}}
\caption{Ablation comparison results for scans of sponge surface. RGB encoded images and en-face projections of (A) original data, (B) stabilized data without overall rotation estimation, (C) stabilized data with only overall rotation estimation and (D) stabilized data of the full proposed algorithm. \DIFadd{The yellow arrows point out rotation artifacts represented by colorful pixels in B-scans, the yellow arrowheads point out colorful pixels caused by the feature changing of the object itself, and the red asterisks mark out aliasing artifacts in en-face projections.}}
\label{fig_abla2}
\end{figure}

 \begin{figure}[tb!]
\centerline{\includegraphics[width=0.95\columnwidth]{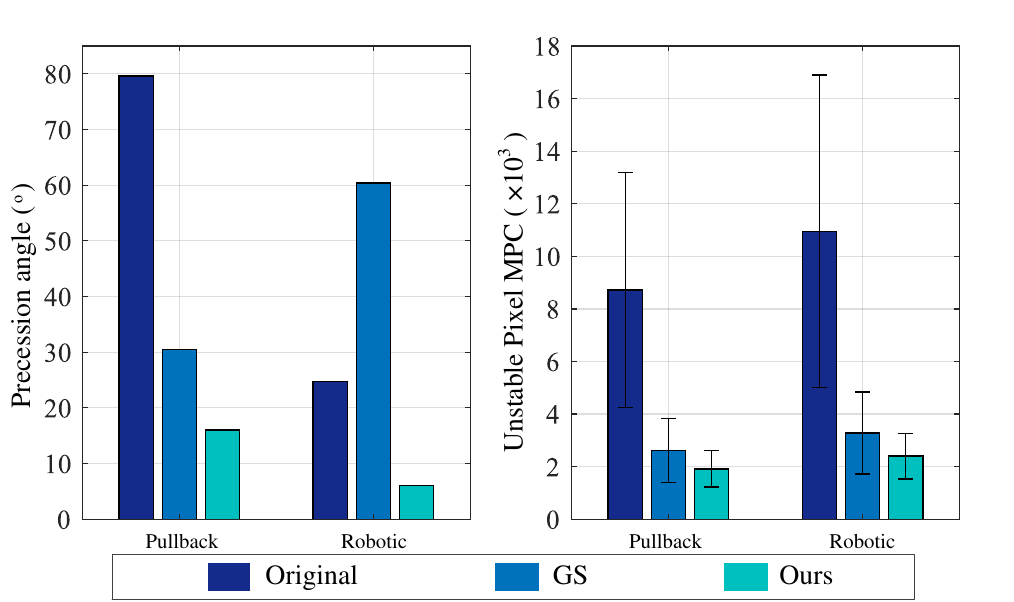}}
\caption{Comparison with GS method. The left figure shows the max precession angles of difference data output, the right figure shows the in-stable pixel MPC comparison.}
\label{fig_gs}
\end{figure}
We use both the conventional internal pullback and the robotic scan and the synthetic video to study the functionality of the two parts of the algorithm (the element-wise \gls{nurd} estimation part, and the overall rotation estimation part). We did a thorough ablation comparison by disabling any of those two parts, and the results are shown in Table \ref{tab_ablatation}. 
The most detrimental modification is to disable the overall estimation part. In this case the proposed algorithm can no longer provide a valid angular value to compensate the drift error, and the precession error is no longer corrected. Sometimes it will even introduce a precession larger than the original precession, for example in the case of \DIFadd{the} real robotic scanning and synthetic videos. 
Disabling the element-wise NURD estimation also affects the reduction of the precession angle, but not as much as disabling the overall estimation. Compared to our full stabilization algorithm, disabling the NURD estimation obviously increases the unstable pixel count, because the overall rotation estimation can not align individual A-lines. This is worse for real scans where the overall rotation may have a larger estimation error due to the sparse features in sheath matching.

\subsubsection{Qualitative results}
Fig. \ref{fig_3d_result} shows qualitative results of pullback scans. To illustrate the NURD instability, we encode each 3 consecutive frames in independent channels of RGB color images\cite{b12}. In this RGB encoding, a colorful part of the image indicates the NURD artifact, and a well stabilized frame sequence should have minimal colorful pixels. We transform the data stream into the en-face projection (see description in section IV.A). All these formats of results are presented in Fig. \ref{fig_3d_result}. It can be seen that the ``colorful" pixels of the original sequence (RGB encoded) is reduced by the proposed algorithm, and without changing the image quality.
The 3D results are consistent with the en-face projections, where the angular rotation is reflected by the shift of high intensity pixels of en-face images. It can be seen that the stabilization algorithm straightens the high intensity lines of the original en-face projection, which maintains the intensity distribution of the whole data stream. \DIFadd{More qualitative visual comparisons can be seen in Supplementary Movies.}

Qualitative results on the ablation study are presented in fig. \ref{fig_abla1} and \ref{fig_abla2}. Keeping only the NURD estimation will introduce iterative distortion to the image, and also limit the precession reduction (fig. \ref{fig_abla1} (B) and fig. \ref{fig_abla2} (B)). Fig. \ref{fig_abla1} (C) and fig. \ref{fig_abla2} (C) show the results with only the overall rotation estimation from the reference data stream. In this case the shaking of the frame sequence can be enlarged due to a large angular estimation error. The fusion stage from the full algorithm can effectively limit these phenomena. 

\begin{table}[!t]
\caption{NURD estimation evaluation }
\label{tab_art}
\centering
\setlength\arrayrulewidth{0.6pt}
\setlength\doublerulesep{0.8pt} 
\begin{tabular}{lC{0.2\columnwidth}C{0.25\columnwidth}C{0.25\columnwidth}}
\toprule[0.8pt] \midrule[0.8pt]
Method              & MSE(Deg$^2$)$\downarrow$& Prec. Reduction (\%) $\uparrow$ & MPC Reduction (\%) $\uparrow$ \\\toprule[0.9pt]
GS \cite{b14}  & 6.47 $\pm$ 1.62  & 12.85   & 70.26 $\pm$ 14.3  \\
Proposed          & \textbf{0.38 $\pm$ 0.09 }  & \textbf{78.86}    &  \textbf{77.78 $\pm$ 8.90} \\ \toprule[1.5pt]
 
\end{tabular}
\end{table}
\subsubsection{Comparison with state of the art}
 In addition, we compare the \gls{nurd} estimation part of the proposed CNN based method and a \DIFadd{state-of-the-art} approach based on the graph path searching (GS)  \cite{b15}, \DIFadd{which is not a learning based method}. Since this online correction method has no drift compensation, we disable the overall rotation estimation to compare the \gls{nurd} estimation performance, using the indirect metrics mean pixel account (MPC) for in-stable pixels of real scans. We also compare them for the synthetic scans, using mean square error (MSE) of \gls{nurd} estimation as a direct metric. Results from original scans and different algorithms are shown in fig. \ref{fig_gs}. Table \ref{tab_art} shows statistics of these comparisons, the \gls{nurd} estimation error of the proposed method is significantly lower than graph searching based method and can better reduce the number of unstable pixels (MPC).

\section{Conclusion}

We have developed a new solution to tackle the rotational distortion problem using deep \gls{cnn}, which can be generalized for scanning situations with different targets and catheters. 
We proposed a new warping vector estimation net to estimate \gls{nurd} between adjacent frames, which has a higher accuracy and robustness compared with the conventional approach in situations where the images have few features. Moreover, we solved the problem of drift error accumulation in iterative video process, with a group rotation estimation net. We were able to apply the \gls{cnn} based algorithm trained on synthetic data to real videos acquired in various scanning conditions. Compared to the A-line level \gls{nurd} vector estimated with tissue information, the orientation estimation using the sheath is less accurate because the feature information provided by the sheath is even more sparse. Nevertheless, this overall rotation is robust and will not be affected by integral drift. The overall rotation is a redundant complementary estimation besides the \gls{nurd} estimation. Fusing the \gls{nurd} warping vector with the overall rotation value can form a robust A-line level warping vector which suppresses the integral drift. 

An ablation study has been performed on different scanning data by calculating a quantitative stabilization metric and qualitative analysis, which shows that both the NURD estimation and overall rotation are essential for the stabilization of the OCT volumetric scanning. The proposed algorithm outperforms other state-of-the-art methods\cite{b15} that iteratively estimate the warping path using a graphic searching approach without a drift compensation module. 

In the literature of OCT stabilization, very little research study can be found, and existing hardware solutions are mainly based on optical marker for calibration \cite{b14}. On the other hand, software methods are for stationary scanning without \DIFadd{longitudinal} translation\cite{b12} or did not consider the precession \cite{b15}. These image based methods will certainly be affected by accumulative error in the volumetric scanning.
The proposed stabilization algorithm is designed for both conventional pullback scan and free 3D scan. For the conventional pullback scan we design a sheath registration and calibration procedure to record reference data. The new proposed 3D scan with robotic tool pullback in a endoscope channel provides two benefits for OCT volumetric imaging. The first is that it reduces the unstable friction and reduces the precession from hardware level. The second benefit is that the sheath in the first frame can be reference for overall rotation estimation, which reduces the procedures of the further stabilization algorithm. 

\DIFadd{The proposed method assumes that the appearance change between consecutive images is more affected by the rotational artifacts than by real changes of tissue. The performance of the algorithm can be impaired in some cases where the appearance of tissue in the field of view changes rapidly, for example when the probe is brought to the tissue very fast or in the presence of peristalsis. However, in the majority of clinical situations, to allow thorough inspection of the tissue and to avoid its damaging, the probe is maneuvered slowly, and thus the algorithm will achieve good stabilization.}


%




\section*{Acknowledgment}
This work was supported by the ATIP-Avenir grant, the ARC Foundation for Cancer research, the University of Strasbourg IdEx, \DIFadd{Plan Investissement d'Avenir and by the ANR (ANR-10-IAHU-02 and ANR-11-LABX-0004-01)} and funded by ATLAS project from the European Union’s Horizon 2020 research and innovation programme under the Marie Sklodowska-Curie grant agreement No 813782. 

\ifCLASSOPTIONcaptionsoff
  \newpage
\fi


\begin{thebibliography}{1}
\bibitem{b_OCT_R} M. Gora, M. Suter, G. Tearney and X. Li, ``Endoscopic optical coherence tomography: technologies and clinical applications [Invited]", \emph{Biomedical Optics Express}, vol. 8, no. 5, pp. 2405-2444, 2017. 
\bibitem{catheter}
M. Atif, H. Ullah, M. Y. Hamza, and M. Ikram, “Catheters for optical coherence tomography,” Laser Physics Letters, 2011. 

\bibitem{b2_c} T. Okamura et al., ``In vivo evaluation of stent strut distribution patterns in the bioabsorbable everolimus-eluting device: an OCT ad hoc analysis of the revision 1.0 and revision 1.1 stent design in the ABSORB clinical trial", \emph{EuroIntervention}, vol. 5, no. 8, pp. 932-938, 2010. 

\bibitem{balloon}
H.-C. Lee, O. O. Ahsen, K. Liang, Z. Wang, C. Cleveland, L. Booth, B. Potsaid, V. Jayaraman, A. E. Cable, H.
Mashimo, R. Langer, G. Traverso, and J. G. Fujimoto, “Circumferential optical coherence tomography
angiography imaging of the swine esophagus using a micromotor balloon catheter,” \emph{Biomed. Opt. Express}, 7(8),
pp. 2927–2942, 2016.
\bibitem{capsule}
 K. Liang, et. al, ``Ultrahigh speed en face OCT capsule for endoscopic imaging", \emph{Biomedical optics express}, 6(4), pp.1146-1163, 2015.
 \bibitem{b1}
Y. Kawase, Y. Suzuki, F. Ikeno, R. Yoneyama, K. Hoshino, H. Q. Ly, G. T. Lau, M. Hayase, A. C. Yeung, R. J. Hajjar, and I.-K. Jang, “Comparison of nonuniform rotational distortion between mechanical IVUS and OCT using a phantom model,” \emph{Ultrasound in Medicine \& Biology}, vol. 33, no. 1, pp. 67–73, 2007. 
\bibitem{b2}
W. Kang, H. Wang, Z. Wang, M. W. Jenkins, G. A. Isenberg, A. Chak, and A. M. Rollins, “Motion artifacts associated with in vivo endoscopic OCT images of the esophagus,” \emph{Optics Express}, vol. 19, no. 21, p. 20722, 2011. 

  


\bibitem{b12} G. V. Soest, J. Bosch, and A. V. D. Steen, “Azimuthal Registration of Image Sequences Affected by Nonuniform Rotation Distortion,” \emph{IEEE Transactions on Information Technology in Biomedicine}, vol. 12, no. 3, pp. 348–355, 2008.

\bibitem{b13} O. O. Ahsen, H.-C. Lee, M. G. Giacomelli, Z. Wang, K. Liang, T.-H. Tsai, B. Potsaid, H. Mashimo, and J. G. Fujimoto, “Correction of rotational distortion for catheter-based en face OCT and OCT angiography,” \emph{Optics Letters}, vol. 39, no. 20, p. 5973-5976, 2014.

\bibitem{b14} N. Uribe-Patarroyo and B. E. Bouma, “Rotational distortion correction in endoscopic optical coherence tomography based on speckle de-correlation,” \emph{Optics Letters}, vol. 40, no. 23, p. 5518-5521, 2015.

 

\bibitem{b15} E. Abouei, A. M. D. Lee, H. Pahlevaninezhad, G. Hohert, M. Cua, P. Lane, S. Lam, and C. Macaulay, “Correction of motion artifacts in endoscopic optical coherence tomography and auto fluorescence images based on azimuthal en face image registration,” \emph{Journal of Biomedical Optics}, vol. 23, no. 01, p. 1-14, 2018.

\bibitem{b16} S. Sathyanarayana, ``Nonuniform rotational distortion (NURD) reduction,'' U.S. Patent 7 024 025 B2, April  4, 2006.

  

\bibitem{b18} G. J. Ughi, “Automatic three-dimensional registration of intravascular optical coherence tomography images,” \emph{Journal of Biomedical Optics}, vol. 17, no. 2, p. 026005, 2012. 

\bibitem{b19} C. Gatta, O. Pujol, O. Leor, J. Ferre, and P. Radeva, “Fast Rigid Registration of Vascular Structures in IVUS Sequences,” \emph{IEEE Transactions on Information Technology in Biomedicine}, vol. 13, no. 6, pp. 1006–1011, 2009.

\bibitem{b20} H. Zhang, W. Yang, H. Yu, H. Zhang, and G.-S. Xia, “Detecting Power Lines in UAV Images with Convolutional Features and Structured Constraints,” \emph{Remote Sensing}, vol. 11, no. 11, p. 1342, 2019.
\bibitem{b_20_a}I. Goodfellow, Y. Bengio, and A. Courville, “Part II: Modern Practical Deep Networks,” in \emph{Deep learning}, Cambridge, MA: MIT Press, 2017. 



\bibitem{b26} J. V. D. Putten, F. V. D. Sommen, M. Struyvenberg, J. D. Groof, W. Curvers, E. Schoon, J. J. Bergman, and P. H. N. D. With, “Tissue segmentation in volumetric laser endomicroscopy data using FusionNet and a domain-specific loss function,” \emph{Medical Imaging 2019: Image Processing}, 2019.

\bibitem{b27} D. Li, J. Wu, Y. He, X. Yao, W. Yuan, D. Chen, H.-C. Park, S. Yu, J. L. Prince, and X. Li, “Parallel deep neural networks for endoscopic OCT image segmentation,” \emph{Biomedical Optics Express}, vol. 10, no. 3, p. 1126, Jul. 2019.


\bibitem{b28} Y. L. Yong, L. K. Tan, R. A. Mclaughlin, K. H. Chee, and Y. M. Liew, “Linear-regression convolutional neural network for fully automated coronary lumen segmentation in intravascular optical coherence tomography,” \emph{Journal of Biomedical Optics}, vol. 22, no. 12, p. 1, 2017.

\bibitem{b29} J. V. D. Putten, M. Struyvenberg, J. D. Groof, T. Scheeve, W. Curvers, E. Schoon, J. J. Bergman, P. H. D. With, and F. V. D. Sommen, “Deep principal dimension encoding for the classification of early neoplasia in Barretts Esophagus with volumetric laser endomicroscopy,” \emph{Computerized Medical Imaging and Graphics}, vol. 80, p. 101701, 2020.

\bibitem{b30} Y. Zeng, S. Xu, W. C. Chapman, S. Li, Z. Alipour, H. Abdelal, D. Chatterjee, M. Mutch, and Q. Zhu, “Real-time colorectal cancer diagnosis using PR-OCT with deep learning,” \emph{Theranostics}, vol. 10, no. 6, pp. 2587–2596, 2020.


\bibitem{b33_bound} K.-K. Maninis, J. Pont-Tuset, P. Arbelaez, and L. V. Gool, “Convolutional Oriented Boundaries: From Image Segmentation to High-Level Tasks,” \emph{IEEE Transactions on Pattern Analysis and Machine Intelligence}, vol. 40, no. 4, pp. 819–833, 2018. 

\bibitem{contour_tradition}
M. Sonka, V. Hlavac, and R. Boyle, Image Processing, Analysis, and
Machine Vision, \emph{2nd ed. Pacific Grove}, CA: PWS, 1999.

\bibitem{contour_DL1}
W. Shen, X. Wang, Y. Wang, X. Bai, and Z. Zhang, “DeepContour: A deep convolutional feature learned by positive-sharing loss for contour detection,” \emph{2015 IEEE Conference on Computer Vision and Pattern Recognition (CVPR)}, 2015. 
\bibitem{contour_DL2}
G. Bertasius, J. Shi, and L. Torresani, “DeepEdge: A multi-scale bifurcated deep network for top-down contour detection,” \emph{2015 IEEE Conference on Computer Vision and Pattern Recognition (CVPR)}, 2015. 
\bibitem{contour_DL3}
J. Yang, B. Price, S. Cohen, H. Lee, and M.-H. Yang, “Object Contour Detection with a Fully Convolutional Encoder-Decoder Network,” \emph{2016 IEEE Conference on Computer Vision and Pattern Recognition (CVPR)}, 2016. 
 
\bibitem{fusion1}
J. L. Crassidis, F. L. Markley, and Y. Cheng, “Survey of Nonlinear Attitude Estimation Methods,” \emph{Journal of Guidance, Control, and Dynamics}, vol. 30, no. 1, pp. 12–28, 2007. 

\bibitem{fusion2} A. Philipp and S. Behnke, ``Robust sensor fusion for robot attitude estimation." \emph{ 2014 IEEE-RAS International Conference on Humanoid Robots}, pp. 218-224., 2014.

\bibitem{fusion3} R. Mahony, T. Hamel and J.M. Pflimlin, ``Complementary filter design on the special orthogonal group SO (3)". \emph{In Proceedings of the 44th IEEE Conference on Decision and Control}, pp. 1477-1484, 2005.

\bibitem{fusion4} 
J. Justa, V. Šmídl, and A. Hamáček, “Fast AHRS Filter for Accelerometer, Magnetometer, and Gyroscope Combination with Separated Sensor Corrections,”
\emph{Sensors}, vol. 20, no. 14, p. 3824, 2020. 

\bibitem{fusion5} 
J. Wu, Z. Zhou, H. Fourati, and Y. Cheng, “A Super Fast Attitude Determination Algorithm for Consumer-Level Accelerometer and Magnetometer,” \emph{IEEE Transactions on Consumer Electronics}, vol. 64, no. 3, pp. 375–381, 2018. 

\bibitem{fusion6} 
D. Gebre-Egziabher, R. Hayward, and J. Powell, “Design Of Multi-sensor Attitude Determination Systems,” \emph{IEEE Transactions on Aerospace and Electronic Systems}, vol. 40, no. 2, pp. 627–649, 2004. 
\bibitem{fusion7} 
Y. S. Suh, “Simple-Structured Quaternion Estimator Separating Inertial and Magnetic Sensor Effects,” \emph{IEEE Transactions on Aerospace and Electronic Systems}, vol. 55, no. 6, pp. 2698–2706, 2019. 
\bibitem{fusion8} 
S. O. H. Madgwick, A. J. L. Harrison, and R. Vaidyanathan, “Estimation of IMU and MARG orientation using a gradient descent algorithm,” z\emph{2011 IEEE International Conference on Rehabilitation Robotics}, 2011. 

\bibitem{stab1} 
J. Yu and R. Ramamoorthi, “Learning Video Stabilization Using Optical Flow,” \emph{2020 IEEE/CVF Conference on Computer Vision and Pattern Recognition (CVPR)}, 2020. 
\bibitem{stab2} 
M. Grundmann, V. Kwatra, and I. Essa, “Auto-directed video stabilization with robust L1 optimal camera paths,” \emph{2011 IEEE/CVF Conference on Computer Vision and Pattern Recognition (CVPR)}, 2011. 
\bibitem{stab3} 
M. Wang, G.-Y. Yang, J.-K. Lin, S.-H. Zhang, A. Shamir, S.-P. Lu, and S.-M. Hu, “Deep Online Video Stabilization With Multi-Grid Warping Transformation Learning,” \emph{IEEE Transactions on Image Processing}, vol. 28, no. 5, pp. 2283–2292, 2019. 

\bibitem{stab4} 
S. Liu, L. Yuan, P. Tan, and J. Sun, “SteadyFlow: Spatially Smooth Optical Flow for Video Stabilization,” \emph{2014 IEEE Conference on Computer Vision and Pattern Recognition}, 2014. 

\bibitem{flow1} 
D. Sun, S. Roth, and M. J. Black, “Secrets of optical flow estimation and their principles,” \emph{2010 IEEE Computer Society Conference on Computer Vision and Pattern Recognition}, 2010. 
\bibitem{flow2} E. Ilg, N. Mayer, T. Saikia, M. Keuper, A. Dosovitskiy, and T. Brox, “FlowNet 2.0: Evolution of Optical Flow Estimation with Deep Networks,” \emph{2017 IEEE Conference on Computer Vision and Pattern Recognition (CVPR)}, pp. 2462-2470, 2017.


\bibitem{dilated} 
F. Yu and V. Koltun, ``Multi-scale context aggregation by dilated convolutions". \emph{Preprint at
arXiv}, https://arxiv.org/abs/1511.07122, 2015.

\bibitem{b37_de_con} M. D. Zeiler, G. W. Taylor, and R. Fergus, ``Adaptive deconvolutional networks for mid and high level feature learning".
\emph{IEEE International Conference on Computer Vision (ICCV)}, pp. 2018–2025, 2011.

\bibitem{b37_up_con} J. Long, E. Shelhamer, and T. Darrell, ``Fully convolutional
networks for semantic segmentation". \emph{Conference on Computer Vision and Pattern Recognition (CVPR)}, 2015.


\bibitem{b22_google} C. Szegedy, W. Liu, Y. Jia, P. Sermanet, S. Reed, D. Anguelov, D. Erhan, V. Vanhoucke, and A. Rabinovich, “Going deeper with convolutions,” \emph{2015 IEEE Conference on Computer Vision and Pattern Recognition (CVPR)}, 2015.
\bibitem{b35_vgg} K. Simonyan and A. Zisserman, “Very deep convolutional networks for large-scale image recognition,” \emph{Int. Conf. Learn. Represent.}, 2015.
\bibitem{b36_resnet} K. He, X. Zhang, S. Ren,  and J. Sun, ``Deep residual learning for image recognition". \emph{Proceedings of the IEEE conference on computer vision and pattern recognition}, pp. 770-778. 2016.
\bibitem{oscar}
O. C. Mora, P. Zanne, L. Zorn, F. Nageotte, N. Zulina, S. Gravelyn, P. Montgomery, M. de Mathelin, B. Dallemagne, and M. J. Gora, “Steerable OCT catheter for real-time assistance during teleoperated endoscopic treatment of colorectal cancer,” \emph{Biomedical Optics Express}, vol. 11, no. 3, p. 1231, 2020. 


\bibitem{stras}
L. Zorn, F. Nageotte, P. Zanne, A. Legner , B. Dallemagne, J. Marescaux, and M. de Mathelin ``A Novel Telemanipulated Robotic Assistant for Surgical Endoscopy: Preclinical Application to ESD," \emph{ IEEE Transactions on Biomedical Engineering},  p. 797-808, Vol. 65, 4, 2018.

\bibitem{b34_relu} A. L. Maas, A. Y. Hannun, and A. Y. Ng, “Rectifier nonlinearities improve neural network acoustic models,” \emph{ICMLWorkshop on Deep Learning for Audio, Speech, and Language Processing (WDLASL
2013)}, vol. 28., 2013.

\bibitem{b41_data1} T. Wang, et al., ``Heartbeat OCT: in vivo intravascular megahertz-optical coherence tomography", \emph{Biomedical optics express}, 6(12), pp.5021-5032. 2015. 

\bibitem{b41_data2} M.J. Gora, et al., ``Tethered capsule endomicroscopy enables less invasive imaging of gastrointestinal tract microstructure". \emph{Nature medicine}, 19(2), pp.238-240. 2013.

\bibitem{b41_data3} S.W. Lee, et al., ``Quantification of airway thickness changes in smoke-inhalation injury using in-vivo 3-D endoscopic frequency-domain optical coherence tomography". \emph{Biomedical optics express}, 2(2), pp.243-254, 2011.

\bibitem{b40pytorch} A. Paszke, et al., ``Automatic
differentiation in pytorch". In NIPS-W, 2017.

\bibitem{b_bn} S. Ioffe and C. Szegedy, `` Batch normalization: Accelerating deep
network training by reducing internal covariate shift". \emph{In International Conference on Machine Learning (ICML)}, 2015.
\bibitem{b_adam} D. P. Kingma and J. Ba, ``Adam: A method for stochastic optimization." \emph{2015 International Conference on Learning Representations}, 2015.
\bibitem{b_start_small}
Y. Bengio, “Global Optimization Strategies,” in \emph{Learning deep architectures for AI}, pp.99 - 104., 2009. 
\bibitem{b_ini} K. He, X. Zhang, S. Ren, and J. Sun, ``Delving deep into rectifiers: Surpassing human-level performance on imagenet classification." \emph{In
IEEE International Conference on Computer Vision (ICCV)},  2015.
\bibitem{b_dpout}G. E. Hinton, N. Srivastava, A. Krizhevsky, I. Sutskever, and R. R. Salakhutdinov, ``Improving neural networks by preventing coadaptation of feature detectors." arXiv:1207.0580, 2012.

\end{thebibliography}
\end{document}